\documentclass[graybox]{svmult}

\usepackage{mathptmx}        
\usepackage{helvet}          
\usepackage{courier}         
\usepackage{type1cm}         
\usepackage{makeidx}         
\usepackage{graphicx}        
\usepackage{multicol}        
\usepackage[bottom]{footmisc}
\usepackage{url}
\usepackage{amsmath,amssymb}
\usepackage{cite, microtype}

\definecolor{red}{rgb}{1,0,0}
\definecolor{green}{rgb}{0,1,0}
\definecolor{blue}{rgb}{0,0,1}

\usepackage{soul}

\makeindex                   
\begin{document}

\title*{The Langevin Approach: 
        a simple stochastic method for complex phenomena}

\titlerunning{The Langevin Approach}

\author{N.~Reinke, A.~Fuchs, W.~Medjroubi, P.G.~Lind, M.~W\"achter and J.~Peinke}

\authorrunning{Reinke, Fuchs et al.}

\institute{Nico Reinke, Andr\'e Fuchs, Wided Medjroubi, Pedro Lind, 
           Matthias W\"achter and Joachim Peinke \at 
           ForWind - Center for Wind Energy Research, Institute of Physics,
           Carl von Ossietzky University of Oldenburg, 26111 Oldenburg, Germany,
           \email{nico.reinke@uni-oldenburg.de}}

\maketitle

\abstract*{%
  We describe a simple stochastic method, so-called Langevin approach,
  which enables one to extract evolution equations of stochastic
  variables from a set of measurements. Our method is parameter-free
  and it is based on the nonlinear Langevin equation. Moreover, it can
  be applied not only to processes in time, but also to processes in
  scale, 
  given that the data available shows
  ergodicity. This chapter introduces the mathematical foundations of
  the Langevin approach and describes how to implement it
  numerically. A specific application of the method is presented,
  namely to a turbulent velocity field measured in the laboratory,
  retrieving the corresponding energy cascade and comparing with the
  results from a computational simulation of that experiment. In
  addition, we describe a physical interpretation bridging between
  processes in time and in scale. Finally, we describe extensions of
  the method for time series reconstruction and applications to other
  fields such as finance, medicine, geophysics and renewable
  energies.}

\abstract{%
  We describe a simple stochastic method, so-called Langevin approach,
  which enables one to extract evolution equations of stochastic
  variables from a set of measurements. Our method is parameter free
  and it is based on the nonlinear Langevin equation. Moreover, it can
  be applied not only to processes in time, but also to processes in
  scale, 
  given that the data available shows
  ergodicity. This chapter introduces the mathematical foundations of
  this Langevin approach and describes how to implement it
  numerically. In addition, we present an application of the method to
  a turbulent velocity field measured in laboratory, retrieving the
  corresponding energy cascade and comparing with the results from a
  computer fluid dynamics (CFD) simulation of that experiment.
  Finally, we describe extensions of the method for time series
  reconstruction and applications to other fields such as finance,
  medicine, geophysics and renewable energies.}
 

\section{Introduction}

"The present state of the universe is an effect of its past states
and causes its future one". Such a claim is a fundamental assumption
in every physical approach to our surrounding nature and was
mathematically defended for the first time two centuries ago, in 1814,
by Simon Laplace.
Laplace had a dream \cite{laplace},
one where "an intellect at a certain moment would know all forces that 
set nature in motion, and all positions of all items of which nature is 
composed, [...] vast enough to submit these data to analysis [...],
to embrace in a single formula [all movements of the universe]".
Why was this a dream? Because there are strong arguments against it, such as
thermodynamic irreversibility, 
quantic indeterminacy and nonlinear
sensitivity to initial states.
But there is also a practical reason: such a high-dimensional
problem, due to its huge number of variables, would only be computable if
one would take 
as model of reality the reality itself, an approach which is pointless.
To model reality one needs simplifications and stochastic methods enables 
one to simplify reality in several adequate ways. 
In this chapter we describe one of such ways, which, in the last fifteen
years, has been successfully applied in several fields \cite{physrepreview}.
\begin{figure}[t]
\center\includegraphics[width=0.95\textwidth]{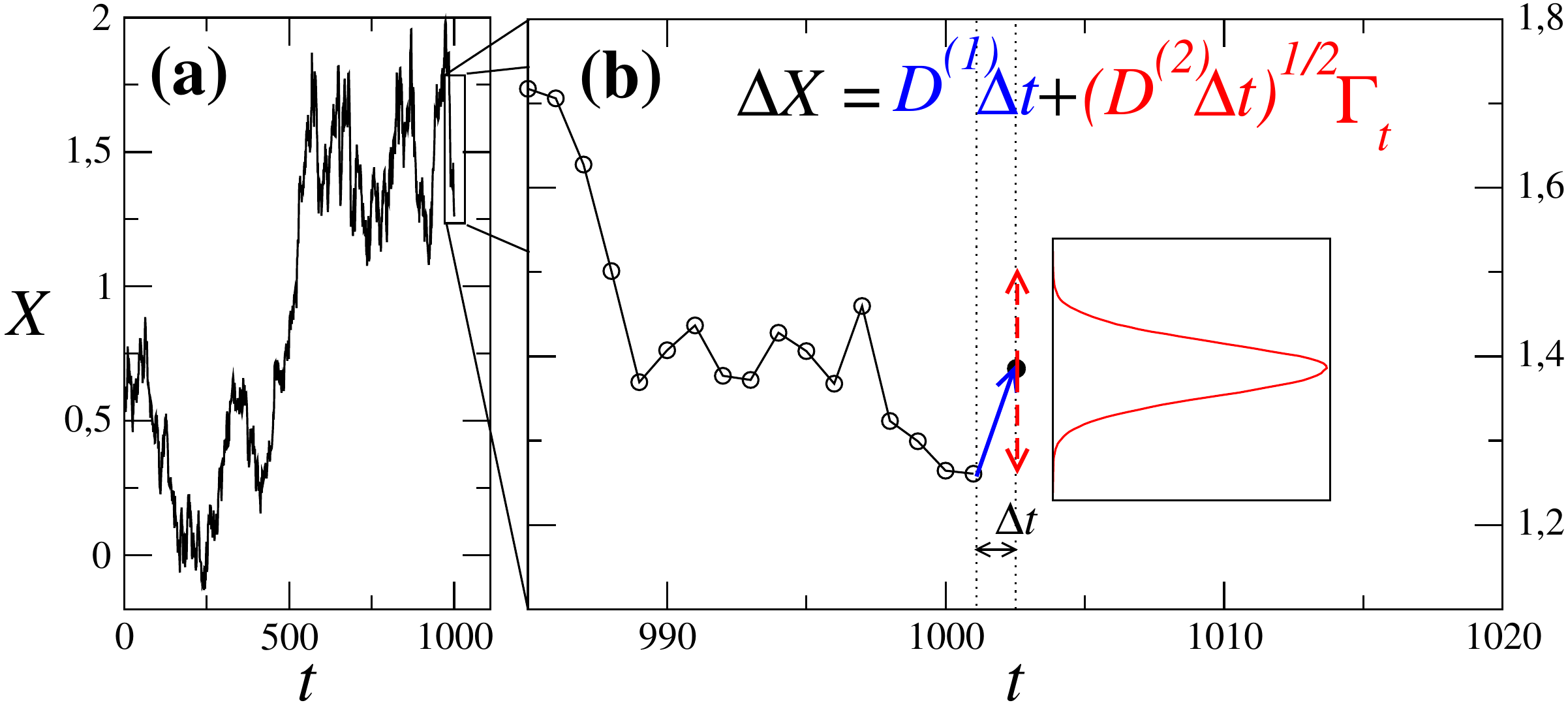}
\caption{\protect
         Illustration of $F$ and $G$ in
         equation (\ref{generalstochequation}). 
         The deterministic contribution, $F=D^{(1)}$, which drives the
         system according to $X\to X+ F\Delta t$ and a stochastic
         contribution $G=\sqrt{D^{(2)}}\Gamma_t$ that is added to it,
         according to some probability distribution.
         Both functions $D^{(1)}$ and $D^{(2)}$ have a well-defined
         meaning and can be extracted from sets of measurements.
         See Sec. ~\ref{sec:langevin}.} 
 \label{fig01}
\end{figure}

As an illustration we address the problem of turbulence, one of the
central open problems in physics \cite{turbulencebook}.
A turbulent fluid is governed by the so-called Navier-Stokes equations
which cannot be approached analytically in all their detail. Therefore
one handles Navier-Stokes equations numerically, developing
discretization schemes which yield the solution of one specific problem.
Such discretization in space and time corresponds in general to
high-dimensional problems which, in the limit of infinitely small 
discrete elements, leads to infinite many degrees of freedom.
Such numerical approaches to the equations governing turbulence enabled
rather successful insight and modeling approaches for fundamental
physics and engineering applications \cite{turbulencebook}.

However, the Navier-Stokes equations, which are purely deterministic, could be
substituted by a stochastic approach, using only a few -- the essential
-- variables, say $X_i$ ($i=1,\dots,N$) and incorporating the rest of
the degrees of freedom in a 
``stochastic bag''. In this way one arrives to evolution equations
of the type:
\begin{equation}
\frac{dX_i}{dt} = F_i(X_1,\dots,X_N,t) +
G_i(X_1,\dots,X_N,\Gamma_1,\dots,\Gamma_M,t)
\label{generalstochequation}
\end{equation}
where function $F_i(X_1,\dots,X_N)$ is a deterministic
function depending on each variable $X_i$ and function
$G_i(X_1,\dots,X_N,\Gamma_1,\dots,\Gamma_M)$ depends not only on 
variables $X_i$ but also on stochastic forces $\Gamma_j$ ($j=1,\dots,M$).
For $G_i\equiv 0$, equation (\ref{generalstochequation}) reduces to a
deterministic dynamical system and for $G_i\sim 0$ one can take it as
a deterministic dynamical system subjected to small noise of constant
amplitude \cite{nontimeseries}.

In general however, not only function $G$ cannot be neglected but
it possesses a much more complicated (nonlinear) dependence on the accounted
variables. Such functional dependence of $G$ is important, for
instance when one intends to describe physical features of a process
underlying a set of data or when aiming at predicting or reconstructing a
set of observations.  

Having properly defined an equation such as
equation (\ref{generalstochequation}), it should be possible to reconstruct 
series of values of one variable, say $X\equiv X_i$ in a
statistical sense, i.e.~it should be possible to derive the
conditional probability: 
\begin{equation}
p(X(t+\Delta t)\vert X(t),X(t-\Delta t),\dots,X(t_0)) 
\label{condprobgeneral}
\end{equation}
for each set of values $X(k)$ with $k=t_0,\dots,t$.
 See sketch in Fig.~\ref{fig01}. 

In this chapter we will describe in detail how to derive an evolution
equation such as equation (\ref{generalstochequation}) from a set
of measurements. 
Our method, so-called Langevin approach, is fully introduced in 
Sec.~\ref{sec:langevin}. In in Sec.~\ref{sec:applic} the Langevin approach is applied to 
turbulence by using the data gained from an experimental study in Sec.~\ref{sec:exp}. Furthermore, the Langevin approach is
applied to a numerical simulation of the experimental study in Sec.~\ref{sec:simul}.
A comparative analysis, discussing the results obtained in the
experiments and simulations is given in Sec.~\ref{sec:analysis}.
Section \ref{sec:conclusions} concludes this chapter, discussing
briefly recent trends in the Langevin approach and
other possible fields and topics where it can be successfully applied.

\section{The Langevin approach}
\label{sec:langevin}

To introduce the Langevin approach, we first explain in
Sec.~\ref{subsec:procscale} what processes in 
scale\footnote{In statistics such processes are also known as branching 
processes.} are and relate them with the usual processes in time. In
Sec.~\ref{subsec:assumptions}, the necessary conditions under which the
Langevin approach is applied are given together with a brief
description of how to verify these conditions in empirical and simulated
data. Section \ref{subsec:fokker-planck} 
describes the derivation of the deterministic and stochastic
contributions with a given set of measurements taken from a 
process in scale.
Finally, in Sec.~\ref{subsec:langevin} 
we derive the stochastic evolution equation describing a process in
scale. A particular example is described, namely the "Brownian
motion" in scale, using the Galton box as an illustration, which
complements the usual Brownian motion described by 
Einstein \cite{einstein1905} and Langevin \cite{Langevin1908}.

\subsection{Processes in scale}
\label{subsec:procscale} 

There is a famous poem by Richardson \cite{turbulencebook} about
turbulence which summarizes his important paper from 1920
\cite{richardson}: "a turbulent fluid is composed by a few big eddies
that decay into smaller eddies, these ones into even smaller eddies,
and so on to viscosity". The energy that feeds the turbulent fluid
enters the system on large scales -- through the largest eddies -- and
travels towards smaller and smaller scales up to a minimum scale where
it leaves the system by means of dissipation. In each of these steps
the energy of the larger eddy is randomly distributed between the
smaller ``child'' eddies. 
Such a pictorial view of turbulent energy traveling through a
hierarchical succession of length scales leads to the concept of the
turbulent energy cascade evolving in the spatial length scale as
the independent variable.

One may ask if it would be possible to extract such energy cascade
from empirical data, e.g.~from a set of velocity measurements at one
specific point of the fluid. In the following we show that indeed it is
possible \cite{friedrich1997,renner2001,physrepreview}.
Describing how a property behaves across an ordered series of
different spatial scales is analogous to the more common description of
the evolution of a property in time, with the important difference
that instead of the time-propagator in equation (\ref{condprobgeneral}) one
has now a ``scale-propagator''. 

Assuming that one has a non-negligible stochastic contribution, the
aim is to derive an equation analogous to equation
(\ref{generalstochequation}) where the spatial scale $r$ plays the
role of time $t$. Since the independent variable $r$ accounts for the
size of some structure, like an eddy, we choose for the dependent variable
the difference of an observable $X$ at two distinct
positions, separated by $r$, namely the \emph{increment}
\begin{equation}
  \Delta X_r(x) = X(x+r)-X(x) 
  \label{eq:IncrementDef}
\end{equation}
with $x$ being a specific location in the system. 
Thus, the scale-propagator describes how this increment -- or
difference -- changes when the distance increases or decreases as follows:
\begin{equation} 
p(\Delta X_{r+\Delta r}\vert \Delta X_{r},\Delta X_{r-\Delta
  r},\dots,\Delta X_{r_0}) .
\label{pscale}
\end{equation} 


Four important remarks are due here.
First, the scale increment $\Delta r$ in equation (\ref{pscale}) can in 
general be positive or negative. In fact, as we will see in the 
next sections, the energy in turbulence flows from the largest
scales, of the size of the system itself, toward the smallest scale 
at which dissipation takes place. Therefore, in turbulence one
considers a scale-propagator as in (\ref{pscale}) with $\Delta r<0$.

Second, one should define a proper metric for the scale $r$. Is the
spatial distance the best choice? Or is there a more appropriate functional of
spatial distances? A process in time evolves according to an iteration
from $t$ to $t+dt$. The same should occur for processes in scale.
However, when "iterating" from one scale  to the "next", 
one iterates in a multiplicative way, i.e., from one scale to the next one 
one multiplies the previous scale by some constant $a$, yielding
a succession of scales $r_n=a^n\to r_{n+1}=a^{n+1}=ar_n$. 
A suitable choice of an additive scale, similar to the additive time iteration,
is the logarithmic scale $\log{r}$, since in this case one has
$\log{r_n}=n\log{a}\to\log{r_{n+1}}=(n+1)\log{a}=\log{r_n}+\log{a}$.
This logarithm scale is of importance to understand the
concept of self-similarity, which is closely related to processes in
scale.
Self-similarity is the property that a phenomenon may manifest, 
by showing invariance under multiplicative changes of scale, as
observed in turbulent flows.
Indeed, following Richardson's poem, eddies are self-similar objects, since
multiplying or dividing their size by a proper {\it scaling factor} we 
obtain an eddy again. 
With the logarithmic scale we ``convert''
the multiplicative changes into additive ones.

Third, when analyzing processes in scale, ideally one would consider a
field of measurements taken simultaneously within a spatially extended
region. What one typically has, contrastingly, is a set of
measurements {\it in time} taken at a particular location. To extract
processes in scale from single time series, one requires the property
being measured to be ergodic: the system should display the
same behaviour averaged either over time or over the space of
all the system's states. In the particular case of a turbulent fluid,
ergodicity reduces to the so-called Taylor hypothesis
\cite{turbulencebook}.

Fourth, while the derivation of a propagator in scale may be
helpful for uncovering phenomena such as the energy cascade
in turbulence, one may also aim to bridge from the derived propagator
in scale to a propagator in time which would enable time-series
reconstruction.
As shown in previous works \cite{Nawroth2010}, our Langevin approach
enables such bridging from scale to time.

Henceforth, we will consider a process in scale, i.e.~a succession of 
increments $\Delta X_{s}$ of a measurable property $X$, with:
\begin{equation}
s=\log{\left( \frac{R_{max}}{r} \right )}
\label{lambda}
\end{equation}
taking values from $s_0=0$ (largest scale $r=R_{max}$) to
$s_L=\log{(R_{max}/R_{min})}$ at the smallest scale $r=R_{min}$.
Notice that, $ds=-dr/r$ and therefore, for $dr<0$ one arrives again 
at a positive scale increment.

\subsection{Necessary conditions: stationarity and the Markov property}
\label{subsec:assumptions}

To apply our method, two important features must be met.
First, the set of measures from which one extracts the succession of
increments in scale must be a stationary process.  
Second, the process in scale must be Markovian.

For the process $X_t$ to be stationary, the corresponding conditional
probability in equation (\ref{condprobgeneral}) should be invariant under
a translation in time, $t\to t+T$, $\forall T$. Numerically such
property cannot be tested in sets of measurements. As an alternative,
one usually divides the set of measurements in $n$ subsets of $N/n\gg 1$
data points and computes the first four centred moments. In case the
centred moments do not vary significantly from one subset to the next
one, the set of measurements is taken as stationary.

The Markov condition of the scale process reads \cite{vankampen}:
\begin{equation} 
p(\Delta X_{s+\Delta s}\vert \Delta X_{s},\Delta X_{s-\Delta
  s},\dots,\Delta X_{s_0}) = p(\Delta X_{s+\Delta s}\vert \Delta 
X_{s}) .
\label{MarkovCond}
\end{equation} 
Notice that, an important consequence of the Markov condition is that
any $n$-point statistics on $X$ can be extracted from the two-point 
statistics \cite{risken} on the increments, 
$p(\Delta X_{s+\Delta s}, \Delta X_{s})$, i.e.~ three-point statistics on $X$. 
The two-point joint distribution of the increments contains all
the information of the scale process.

Equation (\ref{MarkovCond}) tells us that, any conditional probability
distribution from the process conditioned to an arbitrarily large
number of previous observations equals the condition probability
conditioned to the single previous observation solely.
Again, such condition is not possible to ascertain in all its
mathematical detail. A weaker version of equation (\ref{MarkovCond})
suitable for numerical implementation is:
\begin{equation}
p(\Delta X_{s+\Delta s}\vert \Delta X_{s},\Delta X_{s-\Delta
  s}) = p(\Delta X_{s+\Delta s}\vert \Delta
X_{s}) .
\label{MarkovCond-Impl}
\end{equation}
Both conditions in equations (\ref{MarkovCond}) and
(\ref{MarkovCond-Impl}) are equivalent under the physically reasonable
assumption that the dependency of the future state on previous states
decreases monotonically with the time-lag. The equality in equation (\ref{MarkovCond-Impl}) can be qualitatively
verified by plotting contour plots in the range of observed values for 
$\Delta X_{s+\Delta s}$ and $\Delta X_{s}$, and fixing $\Delta
X_{s-\Delta s}=\tilde{X}$. It can also be quantitatively tested
through the Wilcoxon test \cite{wilcoxon}, $\chi^2$-test, or by computing a 
Kullback-Leibler distance between both conditional 
distributions \cite{statisticalbook}.

\subsection{The Fokker-Planck equation for increments}
\label{subsec:fokker-planck}

Once the stationarity of our measures as well as the Markov condition
for their increments are fulfilled, we are able to determine multipoint
statistics for our increments. Since the process is Markovian in scale,
it can be easily proven that for any integer $N$ one has:
\begin{eqnarray} 
p(\Delta X_{s}, \Delta X_{s-\Delta s},\Delta X_{s-2\Delta
  s},\dots,\Delta X_{s-N\Delta s}) = & & \cr
 & & \cr
\left [
\prod_{k=1}^N
\frac{p(\Delta X_{s-(k-1)\Delta s}, \Delta X_{s-k\Delta s})}
{p(\Delta X_{s-k\Delta s})}
\right ]
p(\Delta X_{s-N\Delta s}) & & ,
\end{eqnarray} 
and
\begin{equation}
 p(\Delta X_{s-k\Delta s}) =
\int_{-\infty}^{\infty}
p(\Delta X_{s-(k-1)\Delta s}, \Delta X_{s-k\Delta s})
d\Delta X_{s-(k-1)\Delta s} ,
\end{equation}
for all $k=1,\dots,N$.
Thus, all information of our process is incorporated in the two-point
statistics of the increments. 

It is known that \cite{risken}, the conditional probability
distributions obeys the so-called Kramers-Moyal (KM) equation:  
\begin{equation}
\frac{\partial}{\partial s}  
p(\Delta X_{s}\vert \Delta X_{s_0}) =
\sum_{k=1}^{\infty} \left (
-\frac{\partial}{\partial (\Delta X)}
\right )^k 
D^{(k)}(\Delta X,s)  
p(\Delta X_{s}\vert \Delta X_{s_0}) , 
\label{kramersmoyal}
\end{equation} 
with functions $D^{(k)}$, so-called KM coefficients, 
being defined through conditional
moments $M^{(k)}$ in the limit of small scale increments, namely:

\begin{subequations} 
\begin{eqnarray}  
\label{eq:D_k}
D^{(k)}(\Delta X,s) &=& \lim_{\Delta s \to 0}
                \frac{M^{(k)}(\Delta X,s,\Delta s)}{k!\Delta s} \label{Ds}\\
M^{(k)}(\Delta X,s,\Delta s) &=&
                           \int_{-\infty}^{\infty} \left ( 
                            Y-\Delta X  \right )^k
                           p(Y \vert \Delta X_{s})dY .\label{Ms}
\end{eqnarray}
\end{subequations}
Notice that, from equation (\ref{Ds}), one can see that mathematically each KM coefficient of order $k$, 
apart a multiplicative constant $1/k!$, is the derivative of the conditional moment of the same order $k$.

Numerically, there are two ways for deriving KM coefficients.
One is by computing the conditional moments $M^{(k)}$ for 
a range of observed values of $\Delta X$ and $s$, which is divided in
a certain number of bins, and repeating the computation for 
several values of $\Delta s$. 
In the case where the conditional moments depend linearly on $\Delta s$, at least for the
lower range of values, the KM coefficients are taken
as the slope of the linear interpolation of the corresponding
conditional moment in that range of values. In case such linear dependency is not observed, a second procedure is possible:
one computes at once the entire fraction within the limit in equation (\ref{Ds})
again for a range of observed values of $\Delta X$ and $s$, divided in a proper 
number of bins, but this time one takes the range of smallest values of $\Delta s$ 
and through a linear interpolates infers the projection in the plane 
$\Delta s=0$.

The error for $D^{(k)}$ are just given by the 
linear interpolation of the corresponding conditional moments as 
functions of $\Delta s$.
The errors of each value $M^{(k)}(\Delta X, s)$, necessary for 
computing the errors of the linear interpolation, are given by \cite{lind2010}:

\begin{eqnarray}
\sigma^2_{M^{(k)}}(\Delta X,\Delta s) &=& M^{(2k)}(\Delta X,\Delta s) - \left [ 
     M^{(k)}(\Delta X,\Delta s) \right ]^2  .
\end{eqnarray}

The KM equation (\ref{kramersmoyal}) also holds for the single
probability distribution, since multiplying both sides by $p(\Delta
X_{s_0})$ and integrating in $\Delta X_{s_0}$ yields the
same equation for $p(\Delta X_{s})$. 

An important simplification in equation (\ref{kramersmoyal}) follows if the
fourth KM coefficient vanishes or is sufficiently small compared to the
first two KM coefficients. Such simplification is based on Pawula's
Theorem which states that if $D^{(4)}\equiv 0$ then all coefficients
in equation (\ref{kramersmoyal}) are identically zero except the first two.
Consequently the Kramers-Moyal equation reduces to the so-called
Fokker-Planck equation: 
\begin{equation}
\frac{\partial}{\partial s} 
p(\Delta X_{s}\vert \Delta X_{s_0}) =
\left (
-\frac{\partial}{\partial (\Delta X)}
D^{(1)}(\Delta X,s) 
+\frac{\partial^2}{\partial (\Delta X)^2}
D^{(2)}(\Delta X,s) 
\right )
p(\Delta X_{s}\vert \Delta X_{s_0})  .
\label{fokkerplanck}
\end{equation}

For such differential
equation of the single probability function one can derive
differential equations for the structure functions of the
increments \cite{renner2001}. Uncertainties in equation (\ref{eq:D_k}) can be overcome, namely when
estimating the limit, by considering a subsequent optimization of 
$D^{(1)}$ and $D^{(2)}$. This optimization procedure is based in a cost function
derived from the conditional probability density functions, which 
are deduced from both the experimental data and from Kramers-Moyal 
coefficients directly \cite{Kleinhans,Nawroth}.

\subsection{Langevin processes in scale}
\label{subsec:langevin}

The Fokker-Planck equation (\ref{fokkerplanck}) above describes the
evolution of the conditional probability density function $p(\Delta
X_{s}\vert \Delta X_{s_0})$ for a process in scale which
can be generated by a Langevin equation of the form:
\begin{equation}
\frac{d}{ds}(\Delta X) = 
D^{(1)}(\Delta X,s)+\sqrt{D^{(2)}(\Delta X,s)}\Gamma_{s},
\label{langevinscale}
\end{equation}
where $\Gamma_{s}$ is a $\delta$-correlated noise (in scale
$s$) with $\langle \Gamma_{s} \rangle = 0$ and $\langle
\Gamma_{s} \Gamma_{s^{\prime}} \rangle =
\delta(s-s^{\prime})$.
\begin{figure}[t]
\center\includegraphics[width=0.95\textwidth]{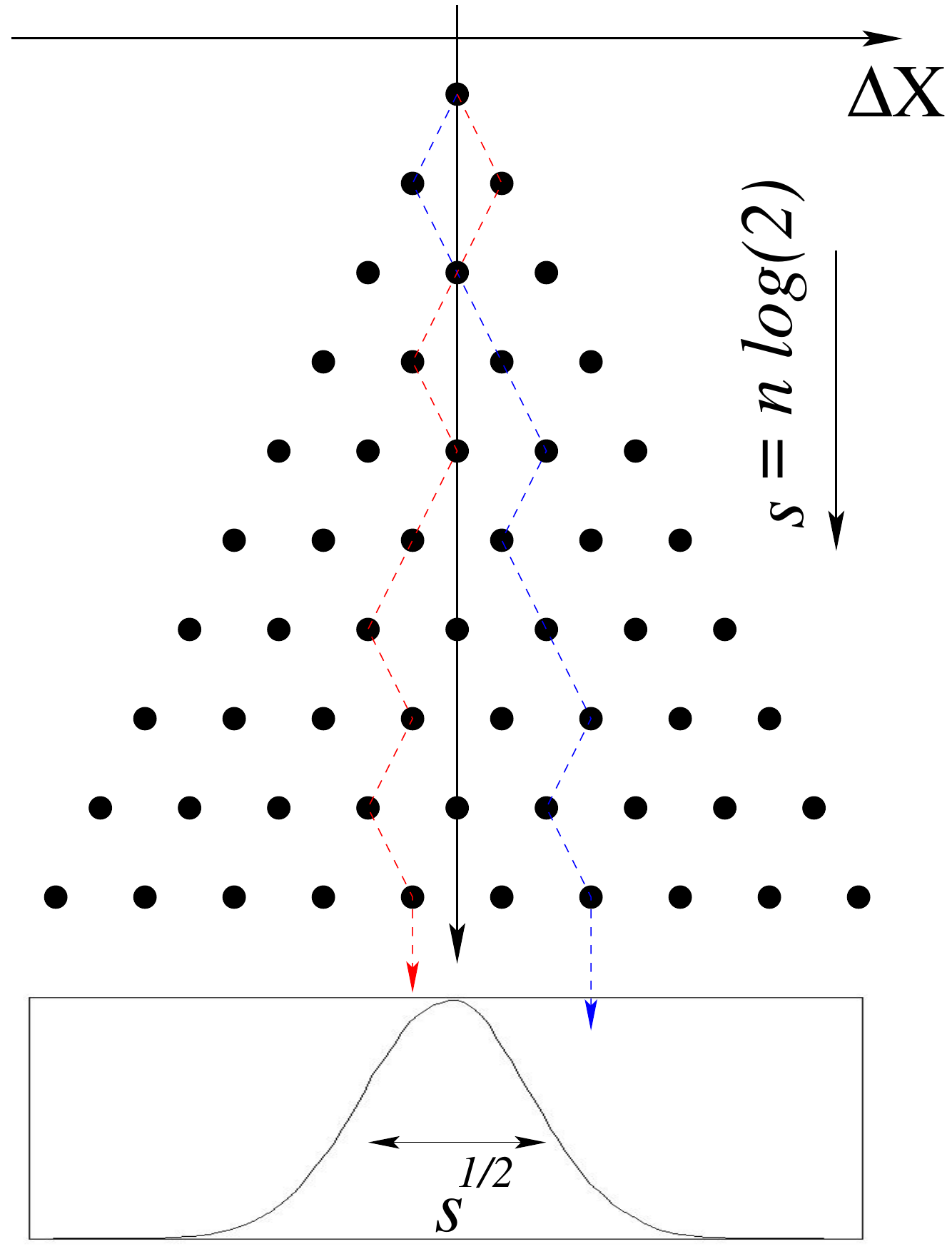}
\caption{\protect
         The Galton box as an illustration of Brownian motion in scale.
         Note that the converging distribution for the increments is
         proportional to $\exp{[-2(\Delta X)^2/s]}$, a Gaussian 
         distribution with standard deviation proportional to 
         $\sqrt{s}$.}
\label{fig02}
\end{figure}

To illustrate the Langevin process in scale described by equation (\ref{langevinscale}),
we consider the particular 
case of $D^{(1)}\propto -\Delta X$ and constant
$D^{(2)}$, reducing the general Langevin equation to the particular
case of Brownian motion "in scale".

What is the Brownian motion {\it in scale}? Though more
abstract than the usual Brownian motion \cite{Langevin1908}, Brownian
motion in scale can be illustrated by a Galton Box \cite{Galton}, as sketched in
figure \ref{fig02}. The Galton box is an apparatus consisting of a vertical board with 
interleaved rows of pins, typically with a constant distance between 
neighbouring pins.  
Balls are dropped from the top, and each time they hit a pin, they 
bounce, left or right, downwards. At the bottom, balls 
are collected in several columns separated from each other.

In a Galton box, the horizontal rows of pins represent the succession of
scales, $s_1,s_2,\dots$, with a constant distance between adjacent
rows, representing the scale increment $\Delta s$. 
From one scale $s_k$ to the next one $s_{k+1}$ the possible ways a
ball can bounce doubles. Since $s$ is in fact a logarithmic scale of
$2^n$, $s_n = n\log{2}$, and therefore $s$ scales linearly with the
vertical distance to the starting point.

As for the horizontal
distance from the centred vertical line, it represents the increments $\Delta X$.
We recall that for processes in scale one has a scale $s$ playing the
role of time and one has increments instead of single values of the
observable $X$.
Thus, similarly to the solution of the original Langevin equation for Brownian
motion, in this case one also obtains a Gaussian distribution of
increment values centred at $\langle \Delta X\rangle =0$
and with a variance proportional to $\Delta s$. Note that the normal
approximation of this binomial distribution is $N(0,s/4)$.

Such illustration of a process in scale is a very simple one.
To properly imagine a picture of general scale processes in turbulence 
two important differences must be considered. First, the energy (i.e. velocity increments) flow from the largest to the smallest scales, which is opposite to the illustration with the Galton box. Second, the KM coefficients for the Galton box are those of the simplest situation that we named as Brownian motion in scale, due to its straightforward parallel with usual Brownian motion. Here the KM coefficients do not depend on scale $s$. In turbulence, as we will see, not only the dependency on the increments is more complicated, but there is an important dependence on the scale $s$.

\section{Applying the Langevin approach to turbulence}
\label{sec:applic}

\subsection{The Langevin approach in laboratory turbulence}
\label{sec:exp}

The experiments were conducted in a closed loop wind tunnel with test section dimensions of
200 cm x 25 cm  x 25 cm (length x width x height) at the University of
Oldenburg. The wind tunnel has a background turbulence intensity of
approximately $2 \%$ for $U_\infty \leq 10$ m/s. 
The inlet velocity was set to 10 m/s, which corresponds to a Reynolds number related to
the biggest grid bar length $L_0$ of about $Re_{L_0}=U_\infty L_0 / \nu =
83800$, where $\nu$ is the kinematic viscosity. Constant temperature anemometry measurements of the velocity 
were performed using (\emph{Dantec 55P01} platinum-plated tungsten
wire) single-hot-wire with a wire sensing length of about $l_w=2.0 \pm
0.1$ mm and a diameter of $d_w = 5$ $\mu m$ which corresponds to a
length-to-diameter ratio of $l_w / d_w \approx 400$. 
A \emph{StreamLine} measurement system by \emph{Dantec} in combination
with CTA Modules 90C10 and the \emph{StreamWare} version 3.50.0.9 was
used for the measurements. The hot-wire was calibrated with
\emph{Dantec Dynamics Hot-Wire Calibrator}. The overheat ratio was set
to 0.8. In the streamwise direction, measurements were performed on
the centerline in the range between 5 cm $\leq x \leq $ 176 cm distance to the
grid. The data was sampled with $f_s= 60$ kHz with a \emph{NI PXI
  1042} AD-converter and 3.6 million samples were collected
per measurement point, representing 60 seconds of measurements
data. To satisfy the Nyquist condition, the data were low-pass
filtered at frequency $f_l=30$ kHz.

\begin{figure}[t]
	\centering
		\includegraphics[height=7cm]{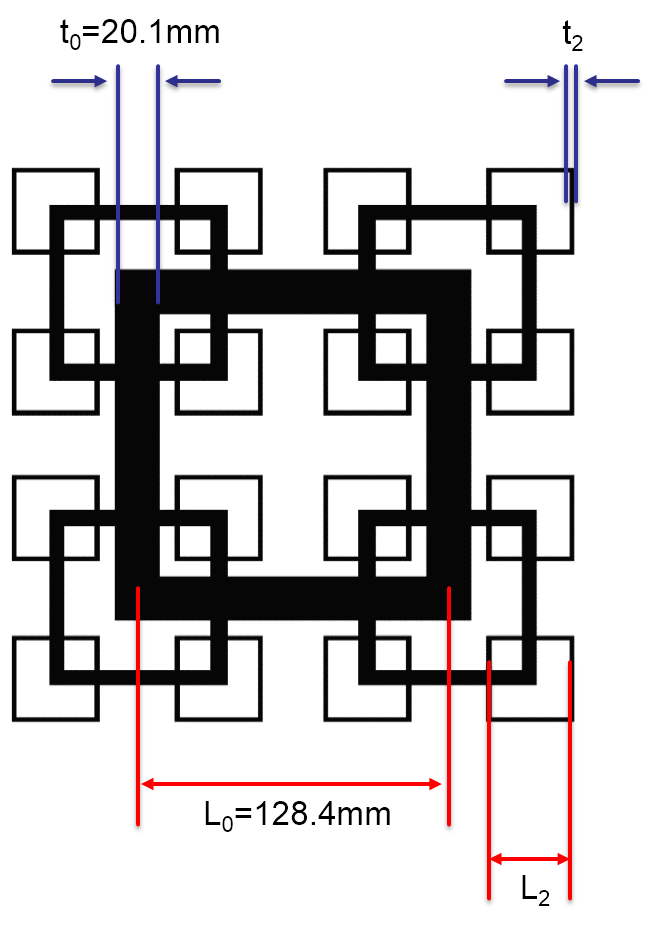}
	\caption{Illustration of the space-filling square fractal grid
          (SFG) geometry, placed at the inlet of the test section for
          the experiments, and considered when implementing the
          corresponding numerical simulations.}
	\label{fig:parameter}
\end{figure}

For the present work, a fractal grid was placed at the inlet of the wind tunnel, see figure \ref{fig:parameter}. In general, fractal grids are constructed from a multiscale collection of obstacles which are based on a single pattern which is repeated in increasingly numerous copies with different scales. The pattern our fractal grid is based on is a square shape with $N=3$ fractal iterations. The fractal iterations parameter is the number the square shape is repeated at different scales. At each iteration $(j=0, ... , N-1)$, the number of squares is four times higher than in the iteration $j-1$. Each scale iteration $j$ is defined by a length $L_j$ and a thickness $t_j$ of the squares bars constituting the grid. The thickness of the square bars in the streamwise direction is constant. The dimensions of the square patterns are related by the ratio of the length of subsequent iterations $R_L = \frac{L_j}{L_{j-1}}$ and by the ratio of the thickness of subsequent iterations $R_t = \frac{t_j}{t_{j-1}}$; respectively. The geometry of the fractal grid we used (also called the space filling fractal grid \cite{Hurst07}) is completely characterized by two further parameters namely
the ratio of the length of the first iteration to the last one $L_r=
\frac{L_0}{L_{N-1}}$ and the ratio of the thickness of the first
iteration to the last one $t_r=\frac{t_0}{t_{N-1}}$.

Contrary to classical grids, fractal grids do not have a well-defined mesh size $M_{eff}$. However, an equivalent effective mesh size was
defined in \cite{Hurst07}.
A complete quantitative description of the N3 fractal grid we used in this study is reported in table \ref{tab:verwendetes_gitter}.
\begin{table}[t] 
\centering
\begin{tabular}[c]{|c|c|c|c|c|c|c|c|c|c|}
  \hline
  $N$ & $\sigma/\%$ & $L_0/mm$ & $t_0/mm$ & $R_L$ & $R_t$ & $L_r$ & $t_r$ & $M_{eff}/mm$ & $T/mm$\\
  \hline
	3 & 36.4 & $128.4$ & $20.1$ & $0.54$ & $0.36$ & $3.5$ & $7.7$ & $24.6$ & $250$ \\
  \hline
\end{tabular}
\caption{Geometrical properties of the utilized fractal grid. $\sigma$ is the blockage ratio and $T$ the cross section of the wind tunnel (and also of the simulation domain).}
\label{tab:verwendetes_gitter}
\end{table}

\subsection{The Langevin approach in simulated turbulence}
\label{sec:simul}

The flow over a fractal grid is described by the three dimensional,
incompressible Navier-Stokes equations. The equations are discretized
and solved using a turbulence model. In this investigation, the
Delayed Detached Eddy Simulation (DDES) \cite{Spalart_Strelets_1997}
with a Spalart-Allmaras background turbulence model
\cite{Spalart_Allmaras_1994}, commonly referred to as SA-DDES is
used. DDES is a hybrid method stemming from the Detached Eddy
Simulation method (DES) \cite{Spalart_Strelets_1997}, which involves
the use of Reynolds Averaged Navier-Stokes Simulation (RANS) at the
wall and Large Eddy Simulation (LES) away from it. This method
combines the simplicity of the RANS formulation and the accuracy of
LES, with the advantage of being less expensive, in terms of
computational time, when compared with pure LES. DDES is an
improvement of the original DES formulation, where the so called
"modelled stress depletion" (or MSD), is treated
\cite{Menter_2004,Spalart_2006}.

The numerical simulation was set up analogous to the experiments in
order to compare the results in a consistent manner. The open source
code OpenFOAM \cite{OpenFOAM:website} was used to solve the
incompressible Navier-Stokes equations. OpenFOAM is based on the finite volume method, and it consists of
a collection of libraries written in C++, which can be used to
simulate a large class of flow problems. For more information about
the available solvers and turbulence models, refer to the official
documentation \cite{OpenFOAM:website}. The solver used in this investigation is the transient solver
pimpleFoam, which is a merging between the PISO (Pressure implicit
with splitting of operator) and SIMPLE (Semi-Implicit Method for
Pressure Linked Equation) algorithms. A second order
central-differencing scheme is used for spatial discretization, and a
backward, second-order time advancing schemes was used. The solver is
parallelized using the Message-Passing Interface (MPI), which is
necessary for problems of this size.  
\begin{figure}[htb]
\center\includegraphics[scale=0.2]{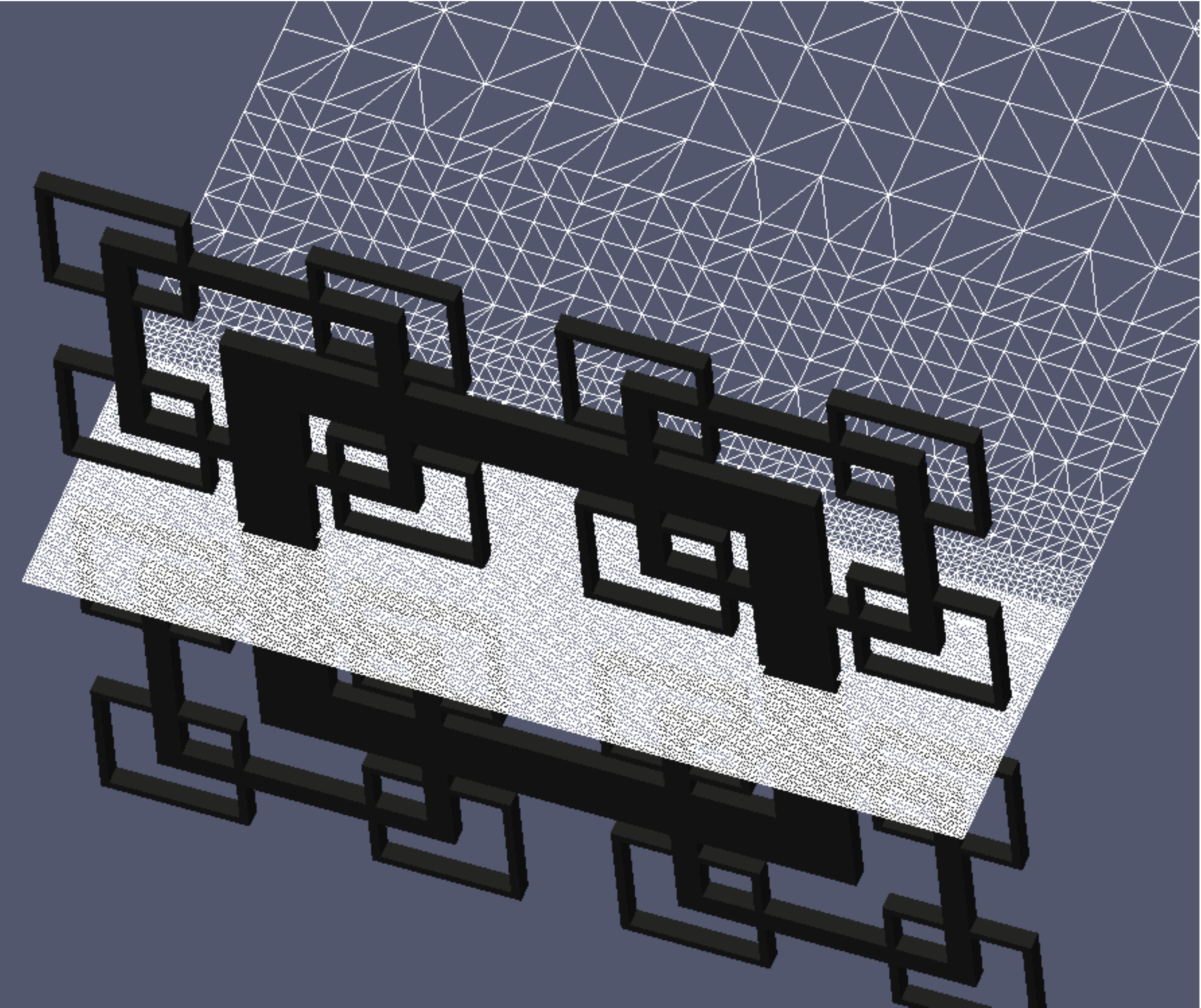}
\caption{Details of the computational mesh used in the computational simulations of the N3 fractal grid.}
\label{fig:N3_fg}
\end{figure}
\begin{figure}[htb]
\center\includegraphics[scale=0.3]{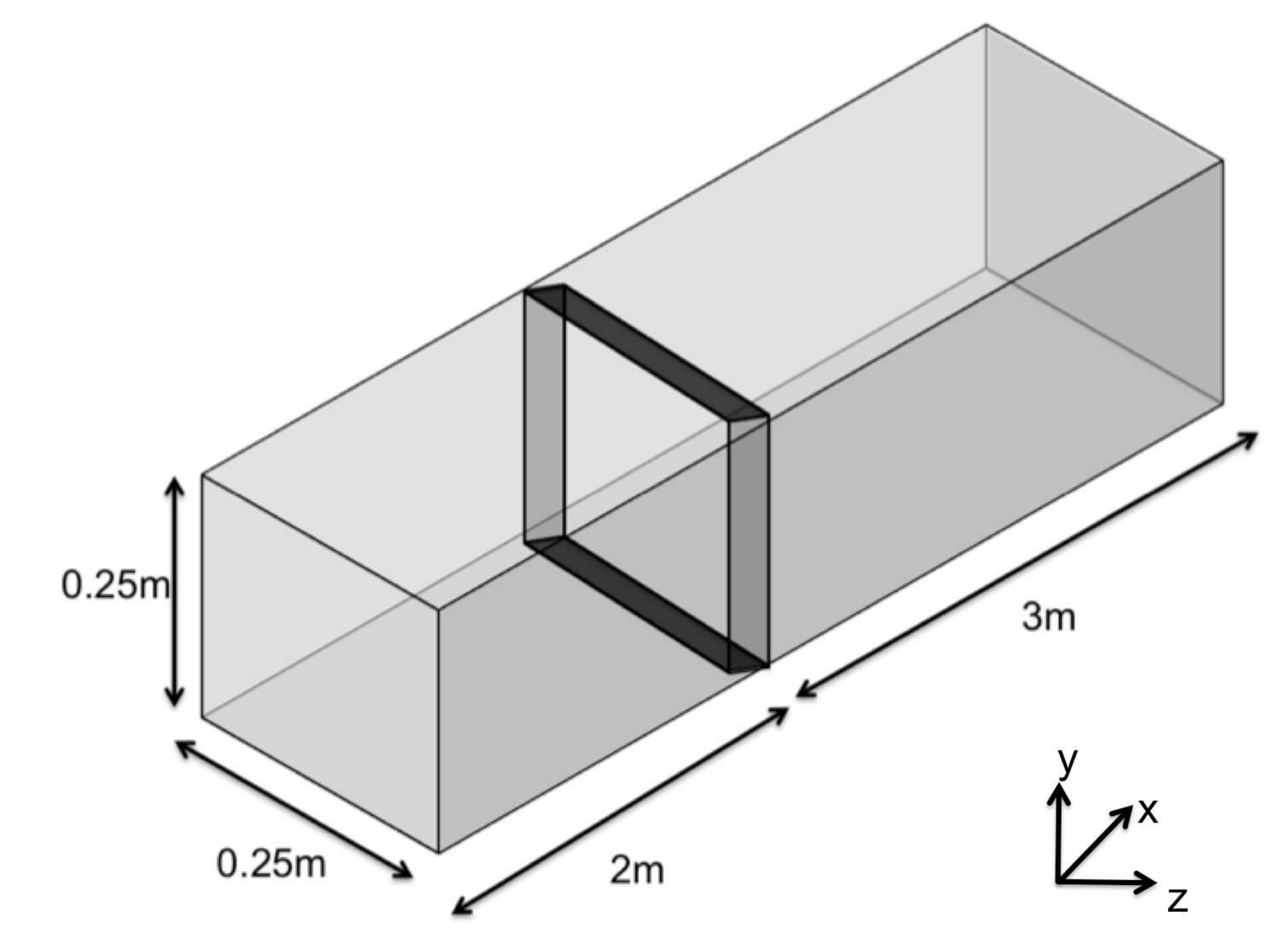}
\caption{Schematic representation of the numerical domain considered and the
  system of coordinates. The fractal grid is positioned where the dark
  block is drawn.} 
\label{fig:N3_fg_domain}
\end{figure}

The numerical mesh was generated using the built-in OpenFOAM meshing
tools blockMesh and snappyHexMesh \cite{OpenFOAM:website}. 
As a result, an unstructured mesh of 24 million cells is obtained,
where regions of interest in the wake are refined, as shown in
Fig. \ref{fig:N3_fg}. The fractal grid is simulated in a domain with similar dimensions as
the real wind tunnel. The domain begins 2 m upstream of the fractal
grid and covers a distance of 2 m downstream (see
Fig. \ref{fig:N3_fg_domain}). The flow-parallel boundaries are treated
as frictionless walls, where the slip boundary condition was applied
for all flow variables. At the inflow boundary, Neumann boundary
condition was used for the pressure, and Dirichlet condition for the
velocity. At the outflow boundary, the pressure was set to be equal to
the static pressure and a Neumann boundary condition was used for the
velocity. On the fractal grid, a wall function is used for the
modified viscosity $\tilde{\nu}$, with the size of the first cell of the mesh in terms of the dimensionless wall distance is $y^{+} \sim 200$ \cite{whiteCFD}. For each simulation, 480 processors were used, and for each time step
4.5 $GB$ of data was collected for post-processing. The data sampling
frequency was 60 $kHz$, chosen to match the experimental one. It took
approximatively 72 hours to simulate one second of data and a total of
20 seconds of numerical data were collected. The data was collected in
the same positions as for the experimental study. The numerical
simulations were conducted on the computer cluster of the ForWind
Group \cite{FLOW01:website}. 

\subsection{Comparative analysis}
\label{sec:analysis}

\begin{table}[b]
\centering
 \begin{tabular}{|c|c|c|c|c|c|} 
 \hline
KM coeff.   &  Data  &          $a$            &        $b$              &       $c$            &         $d$ \\
\hline
$d_{11}$     & \ \  \ \  exp.  \ \ \ \     &  \ \ \ \ $-2.7\times
10^{-5}$   \ \ \ \   &   \ \ \ \  $5.6\times 10^{-4}$   \ \ \ \  &    \ \ \ \  $-0.075$    \ \ \ \     &   \ \ \ \  $-1.0$  \ \ \ \ \\
                   & Sim.    &   $6.7\times 10^{-7}$   & $-7.4\times 10^{-4}$   &   $-0.070$      &   $-0.89$\\
\hline
$d_{22}$     & Exp.    &  $-2.1\times 10^{-6}$   &  $2.0\times 10^{-4}$   &     $-0.0037$    &   $0.059$\\
                   & Sim.    &    $-6.6\times 10^{-6}$ &    $4.1\times
                   10^{-4}$ &     $-5.5\times 10^{-4}$    & $0.072$   \\
\hline
$d_{20}$     & Exp.    &  $0$  &  $0$   &       $0.10$   & $0.18$ \\
                   & Sim.    &    $0$ &  $0$ &       $0.10$   &  $0.21$ \\
\hline
\end{tabular}
\caption{Coefficients $a$, $b$, $c$ and $d$ of the drift and diffusion
  terms for experimental (Exp.) and simulated (Sim.) data, for
  downstream position $x=0.76\hbox{\ m}$.}
\label{table:data}
\end{table}
\begin{figure}[t]
\center
(a)\includegraphics[scale=0.35]{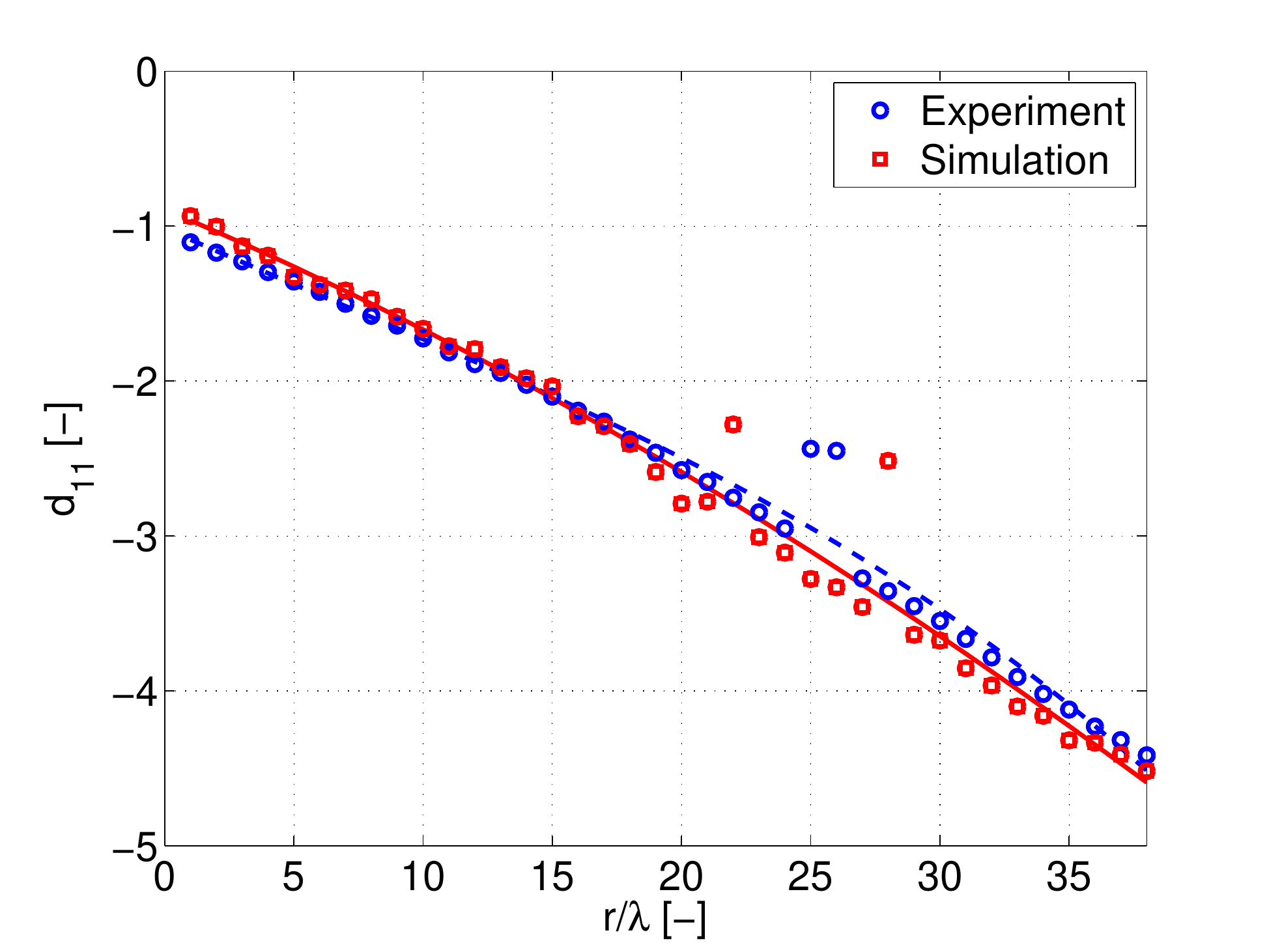}
(b)\includegraphics[scale=0.38]{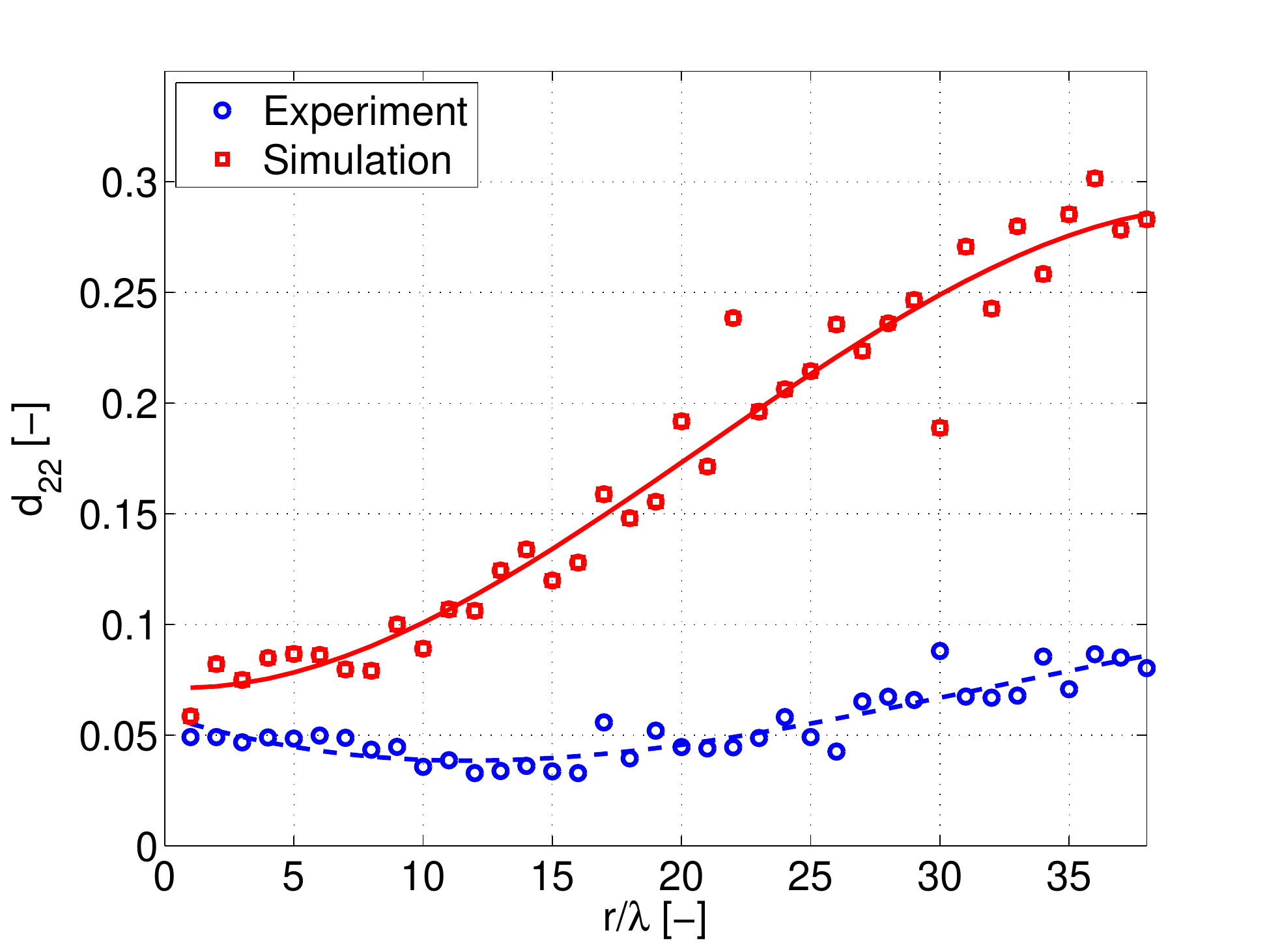}
(c)\includegraphics[scale=0.35]{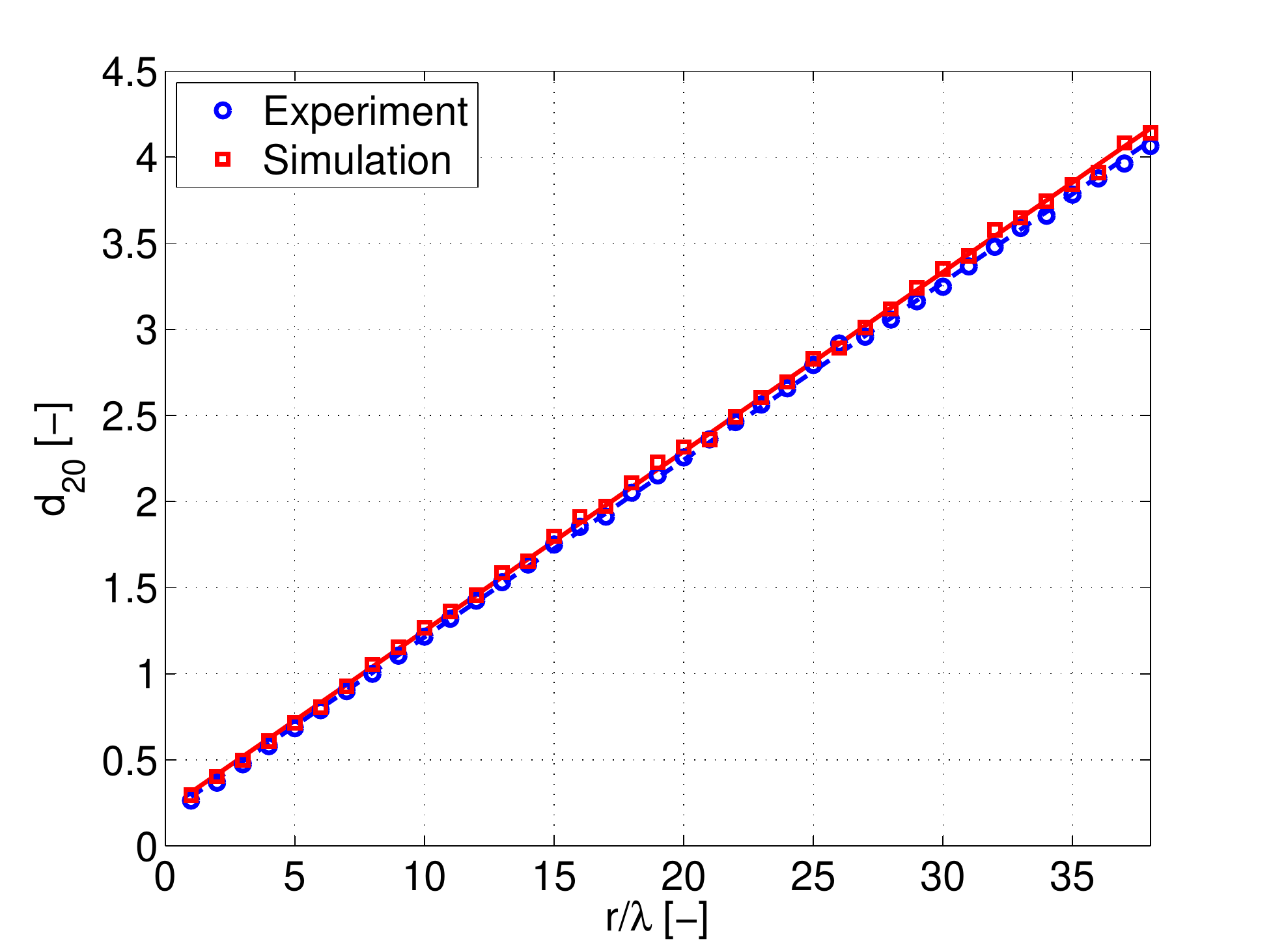}
\caption{Kramers-Moyal coefficients in terms of $d_{11}$, $d_{20}$ and
  $d_{22}$ along the inertial range, $1\le\frac{r}{\lambda}\le L$.}  
\label{fig:KMC}
\end{figure}

We estimate the Kramers-Moyal coefficients $D^{(1,2)}$ for the
experimental and simulated data. The coefficients are commonly
parametrized as follows:
\begin{eqnarray}
D^{(1)}(u,r,x) &=& d_{11}(r,x)\cdot u,\\
D^{(2)}(u,r,x) &=& d_{22}(r,x)\cdot u^2 +  d_{20}(r,x),\\
d_{ii}(r,x) &=&  a_{ii}(x)\left(\frac{r}{\lambda}\right)^3  +
b_{ii}(x)\left(\frac{r}{\lambda}\right)^2 + c_{ii}(x)\frac{r}{\lambda}
+ d_{ii}(x) .
\end{eqnarray}
The results of $D^{(1,2)}$ in terms of $a$, $b$, $c$ and $d$ are shown
in table \ref{table:data}, for the downstream position $x=0.76m$. 
Note that the coefficients strongly depend on the downstream position
$x$. We present and discuss all scales in units of
Taylors microscale $\lambda$.

Figure \ref{fig:KMC} presents Kramers-Moyal coefficients in terms of
$d_{11}$, $d_{20}$ and $d_{22}$ vs. the scale ${r}/{\lambda}$. The
coefficient are calculated within the inertial range. 
We chose this range, because the DDES simulations treat flow structures of
this size as universal and simplify the turbulent flow properties by means of a
sub-grid model, which is in this case the Spalart-Allmaras model. Therefore, a validation with the experimental data within the inertial range is of particular interest. 
Common limits of this region are $\lambda\approx3mm$ (small scale) and
the integral length scale $L\approx12cm$ (large scale). 

The development of the drift term within the inertial range is shown
in \mbox{figure \ref{fig:KMC}(a)}. Comparing the experimental and the
simulated data no significant differences can be observed. Both curves
indicate a stronger drift term with increasing scale, as usual for
turbulent flows. The four outliers are most likely due to the
optimization procedure. Rarely, local minimum are found instate of the
global. 

The development of the diffusion term within the inertial range is
shown in \mbox{figure \ref{fig:KMC}(b)} and (c). 
Figure \ref{fig:KMC}(b) shows the curvature of the diffusion term
$d_{22}$, a very small and sensitive term. Here the experimental and
the simulated data differ in their development. At large scales the
curvature differ significantly ($d_{22,sim}\approx3\cdot
d_{22,exp}$). At small scales the developments draw near, but do not
converge. The magnitude of $d_{22,exp}$ is common, and shows why some
studies neglect $d_{22}$. Figure \ref{fig:KMC}(c) presents the
diffusion term offset $d_{20}$
within the inertial range. 
Such as the drift term, no essential differences between the
experimental and the simulated data can be observed. The linear
increasing of the offset is typical for the inertial range, it
indicates the growth of the increments (velocity difference) or
vortices, respectively,  with scale, cf.~equation (\ref{langevinscale}).

\section{Discussion and Conclusions}
\label{sec:conclusions}

In this chapter we described the so-called Langevin approach, a
stochastic method that enables deriving evolution equations of
stochastic observables, providing important physical insight about the
underlying system.

The method was applied to the problem of turbulence, addressed
experimentally and by means of simulations by extracting the velocity
increment time series, one recorded in a wind tunnel experiment and
one simulated by a delayed detached large eddy simulation (DDES).  
For each case we extracted the functions defining the stochastic
evolution equation, the so-called Kramers-Moyal coefficients, and
parametrized them through polynomials of the scale.

The results show on the one hand good consistency in the two dominating
terms, namely the linear term $d_{11}$ of the first KM coefficient
(drift) and the independent term $d_{20}$ of the second KM coefficient
(diffusion). Other terms, such as the quadratic term $d_{22}$ for the
diffusion, may present deviations that appeal for further
investigation, which will carried out for a forthcoming study focusing
on this specific experiment.

Concerning the Langevin approach as a stochastic method on its own,
three points are worth of mention. First, the method can also be
applied to the usual processes in
time \cite{physrepreview}. For that, one should simply interchange scale
$s$ and increments $\Delta X$ in equations (\ref{kramersmoyal}) and
(\ref{langevinscale}) by time $t$ and observable values $X$
respectively. 

Second, while the method implies the fulfilment of two important
conditions, namely stationarity and markovianity (see
Sec.~\ref{subsec:assumptions}), the method can still be adapted to a
more general situation where one or both this conditions are dropped.
In case the data series is not stationary, the Langevin approach can
be applied to time-windows within which the series can be taken as
stationary \cite{silvio}. 
As a result, one derives a set of KM coefficients as
function of time, one for each time-window.
In the case the data series is not Markovian, for instance due to
measurement (additive) noise an extension is still
possible \cite{lind2010,boettcher2006}. 

Finally, the Langevin approach can be applied to a broad panoply of
different situations in topics ranging technical applications
to biological, geophysical and financial systems, e.g.~electric circuits,
wind energy converters, traffic flow, cosmic microwave background radiation,
granular flows, porous media, heart rhythms, brain diseases such as
Parkinson and epilepsy, meteorological data, seismic time series,
nanocrystalline thin films and biological macromolecules. For a review
on these topics see Ref.~ \cite{physrepreview}.

\section{Acknowledgement}
\label{sec:Acknowledgement}
We gratefully acknowledge the computer time provided by the Facility for Large-Scale Computations in Wind Energy Research (FLOW) of the university of Oldenburg. WM thanks the German Bundesministerium f\"ur Umwelt, Naturschutz und Reaktorsicherheit (BMU), which financed this project. PGL thanks German Federal Ministry of Economic Affairs and Energy (BMWi) as part of the research project 
``OWEA Loads'' under grant number 0325577B.



\bibliography{Bib_TWiSt}

\begin{thebibliography}{10}
\providecommand{\url}[1]{{#1}}
\providecommand{\urlprefix}{URL }
\expandafter\ifx\csname urlstyle\endcsname\relax
  \providecommand{\doi}[1]{DOI~\discretionary{}{}{}#1}\else
  \providecommand{\doi}{DOI~\discretionary{}{}{}\begingroup
  \urlstyle{rm}\Url}\fi

\bibitem{Hurst07}
D.~Hurst, J.V.: Scalings and decay of fractal-generated turbulence.
\newblock Physics of Fluids \textbf{19}, 035,103 (2007)

\bibitem{einstein1905}
Einstein, A.: \"uber die von der molekularkinetischen theorie der w\"arme
  geforderte bewegung von in ruhenden fl\"ussigkeiten suspendierten teilchen.
\newblock Annalen der Physik \textbf{17}, 549--560 (1905)

\bibitem{boettcher2006}
F.Boettcher, J.Peinke, D.Kleinhans, R.Friedrich, P.G.Lind, M.Haase:
  Reconstruction of complex dynamical systems affected by strong measurement
  noise.
\newblock Phys.~Rev.~Lett. \textbf{97}, 090,603 (2006)

\bibitem{FLOW01:website}
Flow01: Facility for large-scale computations in wind energy research (2013).
\newblock \texttt{http://www.fk5.uni-oldenburg.de/57249.html}

\bibitem{friedrich1997}
Friedrich, R., Peinke, J.: Description of a turbulent cascade by a
  fokker-planck equation.
\newblock Phys.~Rev.~Lett. \textbf{78}, 863 (1997)

\bibitem{physrepreview}
Friedrich, R., Peinke, J., Sahimi, M., Tabar, M.: Approaching complexity by
  stochastic methods: From biological systems to turbulence.
\newblock Phys.~Rep. \textbf{506}, 87 (2011)

\bibitem{Galton}
Galton, F.: Natural Inheritance.
\newblock Macmillan (1894)

\bibitem{vankampen}
van Kampen, N.: Stochastic Processes in Physics and Chemistry.
\newblock North-Holland (1999)

\bibitem{Kleinhans}
{Kleinhans}, D., {Friedrich}, R., {Nawroth}, A.P., {Peinke}, J.: An iterative
  procedure for the estimation of drift and diffusion coefficients of langevin
  processes.
\newblock Phys. Letters A \textbf{346}, 42--46 (2005)

\bibitem{Langevin1908}
Langevin, P.: On the theory of brownian motion.
\newblock C.~R.~Acad.~Sci. \textbf{146}, 530--533 (1908)

\bibitem{laplace}
Laplace, P.: A Philosophical Essay on Probabilities.
\newblock Dover Publications 1951 (1814)

\bibitem{lind2010}
Lind, P., Haase, M., Boettcher, F., Peinke, J., Kleinhans, D., Friedrich, R.:
  Extracting strong measurement noise from stochastic series: applications to
  empirical data.
\newblock Phys.~Rev.~E \textbf{81}, 041,125 (2010)

\bibitem{statisticalbook}
Lowry, R.: Concepts and Applications of Inferential Statistics.
\newblock available at http://vassarstats.net/textbook/ (2011)

\bibitem{Menter_2004}
{Menter}, F.R., {Kuntz}, M.: {The Aerodynamics of Heavy Vehicles: Trucks,
  Buses, and Trains}.
\newblock In: Volume 19 of Lecture notes in Applied and Computational Mechanics
  (2004)

\bibitem{Nawroth2010}
{Nawroth}, A.P., {Friedrich}, R., {Peinke}, P.: Multi-scale description and
  prediction of financial time series.
\newblock New Journal of Physics \textbf{12}, 083,021 (2010)

\bibitem{Nawroth}
{Nawroth}, A.P., {Peinke}, J., {Kleinhans}, D., {Friedrich}, R.: Improved
  estimation of fokker-planck equations through optimisation.
\newblock Phys. Rev. E. \textbf{76}(056102) (2007)

\bibitem{OpenFOAM:website}
OpenFOAM: The open source computational fluid dynamics toolbox (2013).
\newblock \texttt{http://www.openfoam.com/}

\bibitem{turbulencebook}
Pope, S.: Turbulence Flows.
\newblock Cambridge Univ.~Press (2000)

\bibitem{renner2001}
Renner, C., Peinke, J., Friedrich, R.: Experimental indications for markov
  properties of small-scale turbulence.
\newblock J.~Fluid Mech. \textbf{433}, 383--409 (2001)

\bibitem{richardson}
Richardson, L.F.: The supply of energy from and to atmospheric eddies.
\newblock Proc.nR.~Soc.~Lond.~A \textbf{97}, 354--376 (1920)

\bibitem{risken}
Risken, H.: The Fokker-Planck Equation.
\newblock Springer, Heidelberg (1984)

\bibitem{silvio}
S.Camargo, Queir\'os, S., C.Anteneodo: Nonparametric segmentation of
  nonstationary time series.
\newblock Phys.~Rev.~E \textbf{84}, 046,702 (2006)

\bibitem{nontimeseries}
Schreiber, T., Kantz, H.: Nonlinear Time-series Analysis.
\newblock Cambridge Univ.~Press, Cambridge (1999)

\bibitem{Spalart_2006}
{Spalart}, P., {Deck}, S., {Shur}, M., {Squires}, K., {Strelets}, M., {Travin},
  A.: A new version of detached-eddy simulation, resistant to ambiguous grid
  densities.
\newblock Theoretical and Computational Fluid Dynamics \textbf{20}(3), 181--195
  (2006)

\bibitem{Spalart_Allmaras_1994}
{Spalart}, P.R., {Allmara}, S.R.: {A One-Equation Turbulence Model for
  Aerodynamic Flows}.
\newblock La Recherche Aerospatiale \textbf{1}, 5--21 (1994)

\bibitem{Spalart_Strelets_1997}
{Spalart}, P.R., {Strelets}, M., {Allmara}, S.R.: {Comments on the feasibility
  of LES for wings, and on a hybrid RANS/LES approach}.
\newblock Advances in DES/LES \textbf{1}, 137--147 (1997)

\bibitem{whiteCFD}
White, F.: Fluid Mechanics.
\newblock McGraw-Hill Higher Education 1998 (1998)

\bibitem{wilcoxon}
Wilcoxon, F.: Individual comparisons by ranking methods.
\newblock Biometrics Bulletin \textbf{1}, 80--83 (1945)

\end{thebibliography}

\end{document}